\documentclass[aps,pra,twocolumn,superscriptaddress,nofootinbib,notitlepage,10pt]{revtex4-1}
\usepackage{epsfig,amssymb,amsmath,mathrsfs,amsfonts,color,amsthm,graphicx,float,color,bm}
\usepackage[colorlinks=true]{hyperref}

\usepackage{graphicx,color}
\usepackage[table]{xcolor}

\usepackage{array}
\newcolumntype{P}[1]{>{\centering\arraybackslash}p{#1}}
\newcolumntype{M}[1]{>{\centering\arraybackslash}m{#1}}

\usepackage{algorithm}
\makeatletter
\renewcommand{\ALG@name}{Protocol}
\makeatother
\usepackage{algpseudocode}
\usepackage[active]{srcltx}
\usepackage{subfigure}
 \def\map#1{\mathcal #1}
\def\d{\operatorname{d}}\def\<{\langle}\def\>{\rangle}
\def\kk{\rangle\!\rangle}
\def\Tr{\operatorname{Tr}}\def\:{\hbox{\bf
    :}}

\def\N{\mathbb N}

\def\grp#1{\mathsf{#1}}

\def\Supp{\mathsf{Supp}}

\def\spc#1{\mathcal{#1}}

\def\set#1{\mathsf{#1}}

\newtheorem{theo}{{Theorem}}
\newtheorem{defi}{{Definition}}
\newtheorem{lemma}{{Lemma}}

\newtheorem{cor}{{Corollary}}

\begin{document}
\title{Optimal universal programming of unitary gates}
\author{Yuxiang Yang}
\email{yangyu@ethz.ch}
\affiliation{Institute for Theoretical Physics, ETH Z\"urich}
\author{Renato Renner}
\email{renner@ethz.ch}
\affiliation{Institute for Theoretical Physics, ETH Z\"urich}
\author{Giulio Chiribella}
\email{giulio@hku.hk}
\affiliation{QICI Quantum Information and Computation Initiative, Department of Computer Science, The University of Hong Kong, Pokfulam Road, Hong Kong}
\affiliation{Department of Computer Science, Parks Road, University of Oxford, Oxford, OX1 3QD, UK}
\affiliation{Perimeter Institute for Theoretical Physics, Waterloo, Ontario N2L 2Y5, Canada}
\affiliation{The University of Hong Kong Shenzhen Institute of Research and Innovation, 5/F, Key Laboratory Platform Building, No.6, Yuexing 2nd Rd., Nanshan, Shenzhen 518057, China}

\begin{abstract} 
A universal quantum processor is a device that takes as input a (quantum) program, containing an encoding of an arbitrary unitary gate, and a (quantum) data register, on which the encoded gate is applied. While no perfect universal quantum processor can exist, approximate processors have been proposed in the past two decades. A fundamental open question is how the size of the smallest quantum program scales with the approximation error. Here we answer the question, by proving a bound on the size of the program and designing a concrete protocol that attains the bound in the asymptotic limit. Our result is based on a connection between optimal programming and the Heisenberg limit of quantum metrology, and establishes an asymptotic equivalence between the tasks of programming, learning, and estimating unitary gates.

%A universal quantum processor is a device that takes as one of its inputs a (quantum) encoding of an arbitrary program, and executes that program on (quantum) data provided as a second input. The well-known No-Programming-Theorem asserts that no perfect universal quantum processor can exist. Nonetheless, approximate programming is still possible, i.e., one can build universal quantum processors that execute a good approximation of the program that is provided as input. Here we answer the question of how the program size scales with the approximation error. We show the achievability of this scaling by designing a concrete protocol, i.e., a prescription of how to determine the program (given a gate) as well as a procedure for how to execute that program. The protocol is based on an gate estimation approach, which operates at the Heisenberg limit of quantum metrology. Our result shows the asymptotic equivalence of three fundamental tasks regarding finite dimensional unitary gates: programming, metrology and learning.  
\end{abstract}

\maketitle
\medskip

\noindent{\em Introduction.} A universal quantum processor is the desideratum of quantum computing. Ideally, one would hope to realise quantum computing in the same way as its classical counterpart, i.e., by inserting data and programs, both in the form of quantum states, into a universal quantum computer. However, the no-programming theorem \cite{nielsen1997programmable} asserts that any universal quantum processor must be approximate, or have a non-zero probability of failure \cite{nielsen1997programmable,hillery2002probabilistic,sedlak2019optimal}. 
% More precisely, it states that any two distinct gates $U_1$ and $U_2$ in an idealised quantum processor must have programs $|\psi_{{\rm P},1}\>$ and $|\psi_{{\rm P},2}\>$ that are orthogonal, i.e., $\<\psi_{{\rm P},2}|\psi_{{\rm P},1}\>=0$. Since there are infinitely many distinct unitary gates, one would need an infinite-dimensional system for the program.

%There are two possible means to circumvent this no-go theorem and to construct approximate universal quantum programmers on systems of bounded dimensions. One is to consider probabilistic programmers \cite{nielsen1997programmable,vidal2002storing,brazier2005probabilistic}. This route has germinated important ideas  like measurement-based quantum computation \cite{raussendorf2003measurement,briegel2009measurement}.
%To build a universal processor on a realistic, finite-dimensional system, one could instead consider approximate universal processors that are allowed to have a small error
It has been shown that approximate universal processors with a finite-size program register do exist  \cite{nielsen1997programmable,kim2001storing,hillery2001programmable,vidal2002storing,brazier2005probabilistic,ishizaka2008asymptotic,kubicki2019resource}. 
There one of the most important questions is to determine the cost-accuracy tradeoff or, more specifically, 
how the program cost, i.e., the number $c_{\rm P}$ of qubits required to store the optimal program, scales with the desired accuracy of implementation, quantified by an approximation error $\epsilon$.
%An answer would be key to the design and the implementation of quantum processors, providing concrete approaches to efficient processors as well as  benchmarks of their performances. 

Over the past two decades, many efforts have been dedicated to finding the optimal approximate universal processor \cite{kim2001storing,hillery2001programmable,ishizaka2008asymptotic,kubicki2019resource} (see also Table \ref{table-compare}). The state-of-the-art result, \cite{kubicki2019resource}, asserts that the optimal program cost $c_{\rm P}$ for a $d$-dimensional unitary quantum gate lies between $c_{\rm low}:=[(1-\epsilon)K]d-(2/3)\log d$ qubits and $c_{\rm upp}:=d^2\log\left(K/\epsilon\right)$ qubits, where $K$ is a universal constant.
Despite all efforts, the precise value for $c_{\rm P}$ remained largely unknown --- especially in the small error regime, where the ratio $c_{\rm upp}/c_{\rm low}$ diverges.

%The optimal program cost $c_{\rm P}\in[c_{\rm low},c_{\rm upp}]$, despite all the effort, remains unidentified. In particular, in the crucial small error region, the ratio $c_{\rm upp}/c_{\rm low}$ diverges.

In this Letter, we close this gap by identifying the optimal scaling of the program cost with the accuracy and therefore solving a long-standing open problem of optimal quantum programming.
Specifically, our program cost scales as $[(d^2-1)/2]\log\left(1/\epsilon\right)$ in the small $\epsilon$ regime, which reduces the cost of the best existing protocol (see $c_{\rm upp}$ above) by half.
The optimal scaling is achieved with a gate learning protocol, where the program is prepared by sending a quantum state through $n$ instances of the gate to learn it \cite{bisio2010optimal}. The gate information is later read out by measuring the program. Our protocol achieves a diamond norm error scaling of $1/n^2$ -- well-known as the Heisenberg limit of quantum metrology \cite{buvzek1999optimal,chiribella2004efficient,bagan2004quantum,hayashi2006parallel}. We thus prove the asymptotic equivalence of quantum gate programming, metrology, and learning. 
%Our result also hints the existence of a universality property that governs asymptotic quantum information processing tasks, shedding light on how the quantum-to-classical transition takes place in the macroscopic limit.

\begin{table*}
\rowcolors{2}{red!10}{white}
\begin{ruledtabular}
    \begin{tabular}{   M{1in}| M{2.95in} | M{2.95in} }
   & Upper bounds & Lower bounds\\  \hline
 Previous works  &  \qquad\quad$d^2\log\left(K/\epsilon\right)$ \cite{kubicki2019resource} \newline \quad\newline $4d^2\log d/\epsilon^2$ \cite{ishizaka2008asymptotic,beigi2011simplified,christandl2018asymptotic} &  $[(1-\epsilon)K]d-(2/3)\log d$  \cite{kubicki2019resource} \newline\qquad \newline  $\log\left(d^2/\epsilon\right)$ \cite{majenz2018entropy} \newline\quad\newline \qquad $\left(\frac{d+1}{2}\right)\log\left(1/d\right)+\left(\frac{d-1}{2}\right)\log\left(1/\epsilon\right)$ \cite{perez2006optimality}\\  \hline
This work & $\left(\frac{d^2-1}{2}\right)\log\left(\Theta(d^2)/\epsilon\right)$  & $\alpha\log\left(\Theta(d^{-4})/\epsilon\right)$ \newline for any $\alpha<(d^2-1)/2$ and sufficiently small $\epsilon$
\end{tabular}\caption{{\bf Comparison of bounds on universal quantum gate programming.} In the table we compare our results on the programming cost with the best previous results (summarised from Table I of Ref.\ \cite{kubicki2019resource}). In the vanishing error regime $\epsilon\to0$, both our lower bound and our upper bound are tighter than all previous results, for the first time closing the gap between the lower and upper bound in this regime. The cost is defined as the number of qubits in the program and the error is evaluated in terms of the diamond norm (\ref{error-diamond}). $K$ denotes a universal constant.  }\label{table-compare}
\end{ruledtabular}
\end{table*}

\begin{figure}  [tb]
\begin{center}
  \includegraphics[width=\linewidth]{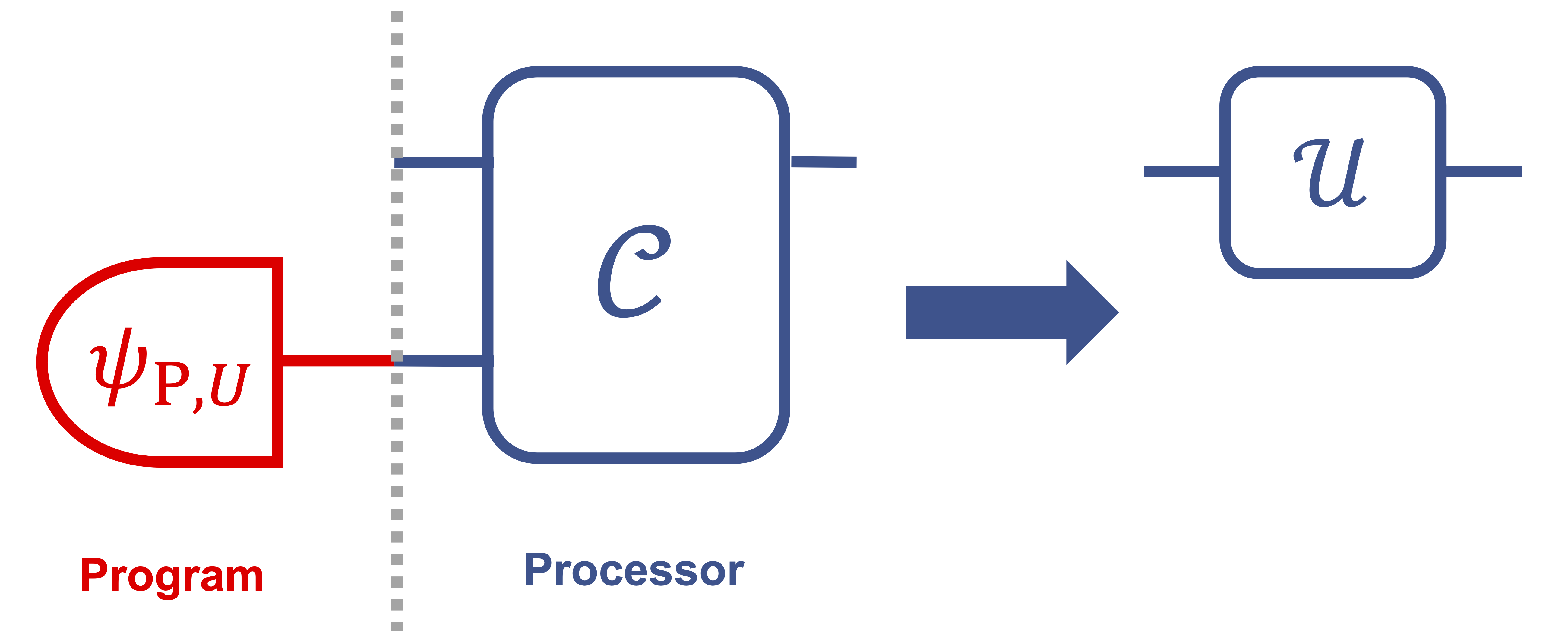}
  \end{center}
\caption{\label{fig-upqp}
  {\bf An approximate universal quantum processor.} An approximate universal quantum processor executes a unitary gate $U$ on a system. It works by plugging a quantum state -- the program for $U$ -- into the processor, which performs a quantum channel $\map{C}$ that approximates $U$ on the system.}
\end{figure}

\medskip

\noindent{\it Preliminaries.} 
We consider programming unitary gates of a system with a $d$-dimensional Hilbert space $\spc{H}$. The gates, up to an irrelevant global phase, form the special unitary group $\grp{SU(d)}$.  
For a pure state $|\psi\>$, we abbreviate its density matrix $|\psi\>\<\psi|$ by $\psi$. Similarity, $\map{U}(\cdot):=U(\cdot)U^\dag$ denotes a unitary channel.

We will use the big-$\Omega$ notation, the big-$O$ notation, and the big-$\Theta$ notation to characterise the asymptotic behaviour of functions. For two non-negative functions $f(n)$ and $g(n)$, we write $f(n)=\Omega(g(n))$ if there exists a constant $c_1>0$ so that $f(n)\ge c_1\,g(n)$ for large enough $n$, $f(n)=O(g(n))$ if there exists a constant $c_2>0$ so that $f(n)\le c_2g(n)$ for large enough $n$, and $f(n)=\Theta(g(n))$ if $f(n)=\Omega(g(n))$ and $f(n)=O(g(n))$. We will also abbreviate $\log_2$ by $\log$.

\medskip

\noindent{\it Approximate universal processors.} A universal quantum  processor consists of two key elements: a family of programs $\{\psi_{{\rm P},U}\}_{U\in\grp{SU(d)}}$, which are quantum states in $\spc{H}_{\rm P}$, and the action of the processor $\map{C}$, which is a quantum channel (i.e.\ a completely positive trace-preserving linear map) acting on the composite Hilbert space $\spc{H}_{\rm S}\otimes\spc{H}_{\rm P}$ of the system and the program. Notice that all information on $U$ should come from the program, and $\map{C}$ must be independent of $U$.
The program cost $c_{\rm P}$ is defined as $\log_2 d_{\rm P}$, with the program dimension $d_{\rm P}$ being the dimension of $\Supp\{\psi_{{\rm P},U}\}_{U\in\grp{SU(d)}}$.

As shown in Figure \ref{fig-upqp}, to run any arbitrary unitary $U$ on the system, one selects the corresponding program $\psi_{{\rm P},U}$ and plugs it into the processor, resulting in the following channel on the system:
\begin{align}
\map{E}_U(\cdot):=\Tr_{\rm P}\left[\map{C}(\cdot\otimes\psi_{{\rm P},U})\right].
\end{align}
A pair $(\map{C},\{\psi_{{\rm P},U}\}_{U\in\grp{SU(d)}})$ is called a \emph{$\epsilon$-universal processor}, if
\begin{align}\label{error-diamond}
\frac12\left\|\map{U}-\map{E}_U\right\|_{\diamond}\le \epsilon\quad\forall\,U\in\grp{SU(d)}.
\end{align}
Here $\|\cdot\|_{\diamond}$ denotes the \emph{diamond norm} \cite{kitaev1997quantum}, which equals the maximum trace distance between the outputs of the two channels, maximized over all input states and over all possible reference systems.

The no-programming theorem \cite{nielsen1997programmable} rules out perfect (i.e.\ $\epsilon=0$) universal processors with finite cost $c_{\rm P}  <\infty$.
This impossibility result raised the question: ``Given a desired accuracy $1/\epsilon$, how big does the program need to be?" This question can of course be subdivided into two, namely to find upper and lower bounds on the program cost $c_{\rm P}$. We summarise the best known results in Table~\ref{table-compare}. Here we are providing both a new lower and a new upper bound, which match in terms of their asymptotic dependence on $1/\epsilon$.

%Focus thereafter has been turned to the requirement of programming (see Table \ref{table-compare} for a summary of the best known results): Given desired accuracy $1/\epsilon$, how big the program has to be? This question can be answered in two directions. One is to determine a lower bound on the program dimension $d_{\rm P}$, defined to be the dimension of $\Supp\{\psi_{{\rm P},U}\}_{U\in\grp{SU(d)}}$. The other is to construct a concrete programmer and see the dependence of its program dimension on $1/\epsilon$.  In the following, we are going to show both a new lower bound and a new upper bound, which finally match each other.

\medskip
\noindent{\em Lower bound on the program cost.}
%We start from the first direction and show a new lower bound on the program dimension. 
%For this purpose, we use a fundamental property of reversible quantum gates, which was first used by us to determine the energy requirement of quantum processors \cite{chiribella2019energy} and further exploited here: 
We first establish a lower bound on the program cost. For this purpose, we exploit an alternative proof of the no-programming theorem, originally developed in the framework of general probabilistic theories \cite{chiribella2010probabilistic}. The idea is that the exact implementation of a unitary gate requires the channel $\map{C}$ to leave the system and the program uncorrelated. Using this fact, the program can be recycled, thereby generating multiple copies of the desired unitary gate. The approximate version of this argument was first used by us to determine the energy requirement of quantum processors \cite{chiribella2019energy} and is further exploited here.

To approximate a unitary quantum gate $U$ with good precision, there should be almost no correlation between the system and the program after we apply $\map{C}$. This means that the complementary channel of $\map{E}_U$, defined as $\overline{\map{E}}_{\rho_{\rm S}}(\cdot):=\Tr_{\rm S}\left[\map{C}\left(\rho_{\rm S}\otimes(\cdot)\right)\right]$, is almost independent of $\rho_{\rm S}$. It further suggests that, instead of discarding the program after one usage, we can \emph{recycle} it:
We can invert the action of $\overline{\map{E}}_{\rho_{\rm S}}$ on the program state by a $(\rho_{\rm S})$-independent operation and get back the original program. The program can be further used, generating multiple uses of $U$ at the cost of an increased approximation error. 
Notice that the argument does not hold for noisy or classical processes. For instance, using a controlled unitary $|0\>\<0|\otimes I+|1\>\<1|\otimes \sigma_z$ and an ancillary qubit $(1/\sqrt{2})(|0\>+|1\>)$ one can (perfectly) implement the channel $\rho\to(1/2)(\rho+\sigma_z\rho\sigma_z)$. However, the system and the ancillary qubit become strongly correlated after the implementation.

By the above argument, we can show (see Appendix for details) that an $\epsilon$-universal processor for a single use of $U$ can be turned into a $(4m\sqrt{2\epsilon})$-universal processor for $m$ uses of $U$ for any $m\ge1$. 
This requires the original program to contain enough information for programming up to $1/\sqrt{\epsilon}$ uses of $U$.
% and it also has to be large enough in dimension.
This fact, in turn, implies a bound on its minimum information content and therefore its size.
 This ultimately leads to the following theorem, which can be regarded as a quantitative version of the no-programming theorem \cite{nielsen1997programmable}:
\begin{theo}[Approximate no-programming theorem]\label{theo-converse}
Consider any $\epsilon$-universal processor with program cost $c_{\rm P}$. 
For any ($\epsilon$-independent) parameter $\delta>0$, the program cost is lower bounded as
\begin{align}
c_{\rm P}&\ge (1-\delta-4\sqrt{2\epsilon})(d^2-1)\log\left(\frac{\delta}{4\sqrt{2\epsilon}(d^2-1)}\right)-1.
\end{align}
%Here the big-O notation is used in the asymptotic limit of $\epsilon\to 0$. 
\end{theo}
This immediately implies the expression for the lower bound stated in Table~\ref{table-compare}.
The key message from the above theorem is that, for any $\alpha<(d^2-1)/2$, the program dimension $d_{\rm P}=2^{c_{\rm P}}$ satisfies
\begin{align}\label{bound-program}
d_{\rm P}&=\Omega\left(1/\epsilon^\alpha\right)
\end{align}
Taking $\epsilon\to0$ in Eq.\ (\ref{bound-program}), one gets $d_{\rm P}\to\infty$, recovering the original no-programming theorem \cite{nielsen1997programmable}.
%Theorem \ref{theo-converse} is an immediate consequence of a more explicit non-asymptotic bound, which we prove in Appendix \ref{app-thm-converse}. It asserts that, 

\medskip

\begin{figure}  [t!]
\begin{center}
  \includegraphics[width=\linewidth]{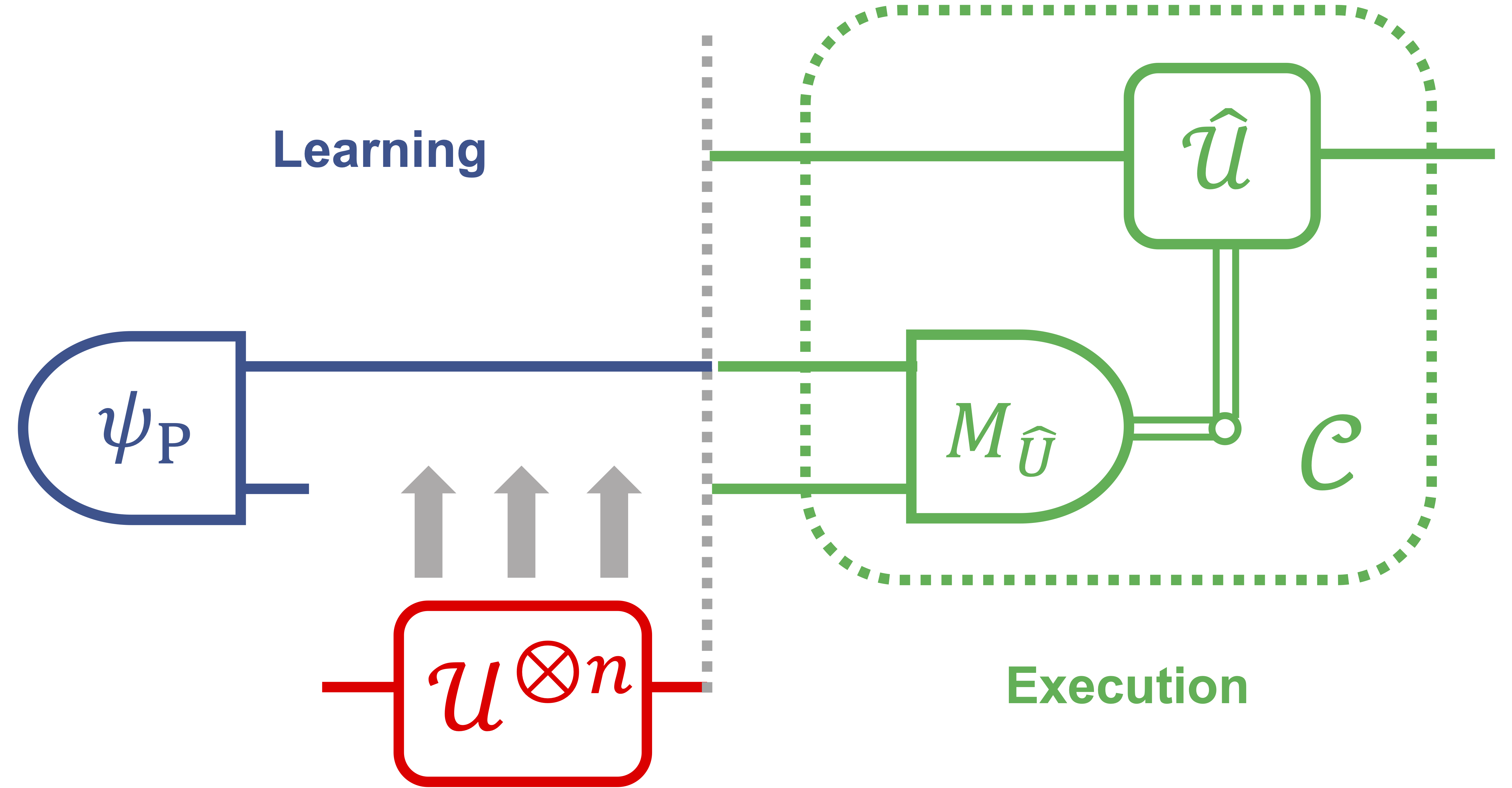}
  \end{center}
\caption{\label{fig-mo}
  {\bf A learning protocol for unitary gates.} In the learning phase, a probe state $\psi_{\rm P}$, possibly entangled with a reference system, is prepared. It is then sent through $n$ parallel instances of $U$, resulting in a program $\psi_{{\rm P},U}$. The program is later measured, and the gate corresponding to the measurement outcome $\hat{U}$ is performed on the system.}
\end{figure}

\noindent{\it Optimal approximate universal processor.} Next we construct an approximate universal processor that achieves the bound in Theorem~\ref{theo-converse}. 
Our processor works in a measure-and-operate (MO) fashion, as illustrated in Figure \ref{fig-mo}. It measures the input program $\psi_{{\rm P},U}$ with a suitable POVM $\{\d\hat{U}\,M_{\hat{U}}\}_{\hat{U}\in\grp{SU(d)}}$, where $\d\hat{U}$ is the Haar measure. The measurement yields an estimate $\hat{U}$ of the gate $U$, and the processor performs the corresponding gate on the system. Explicitly, our optimal processor obeys the following procedure:
\begin{algorithm}[H]
  \caption{A MO universal processor.}
  \label{alg-MO-program}
   \begin{algorithmic}[1]
  % \setcounterref{ALC@line}{0} 
   \State (Generating the program.)\newline  Apply $U^{\otimes n}$ to a suitable quantum state $|\psi_{{\rm P}}\>$.
   \State Measure $|\psi_{{\rm P},U}\>:=U^{\otimes n}|\psi_{{\rm P}}\>$ with $\{\d\hat{U}\,M_{\hat{U}}\}_{\hat{U}\in\grp{SU(d)}}$.
   \State Apply $\hat{U}$ to the state of the system, where $\hat{U}$ is the measurement outcome.
   \end{algorithmic}
\end{algorithm}

%{\color{red} I found it hard to say much about the concrete form of the program. One thing, however, that we may say is that our programs are more orthogonal for more distinguishable unitaries, matching the intuition in \cite{nielsen1997programmable}.}

The program in Protocol \ref{alg-MO-program} is prepared by applying $n$ parallel uses of $U$ on a quantum state (called the \emph{probe state}). The performance of this processor is then determined jointly by the choice of the probe state and the choice of the POVM $\{\d\hat{U}\,M_{\hat{U}}\}_{\hat{U}\in\grp{SU(d)}}$. 
It is known from quantum metrology \cite{chiribella2004efficient,bagan2004quantum,kahn2007fast} that the performance of the measurement is optimised using non-product probe states and POVMs. %\footnote{Here we take a Bayesian perspective of quantum metrology, where the gate is assumed to be drawn randomly according to a uniform measure. This is to be distinguished from the minimax perspective (see, e.g., Refs.~\cite{giovannetti2006quantum,giovannetti2011advances}) which is also very common.} 
In Appendix, we identify a probe state and a POVM which, when incorporated into Protocol~\ref{alg-MO-program}, yields an optimal processor asymptotically achieving the $({(d^2-1)/2})\log(1/\epsilon)$ scaling bound of Theorem~\ref{theo-converse}.
\begin{theo}\label{theo-opt-tomography}
Consider the estimation of an unknown unitary gate on a $d$-dimensional quantum system. When $n\ge 2 d (d-1)$ uses of the gate are available, the diamond norm error
 for the optimal  estimation is bounded as
\begin{align}\label{F-wc-tomo}
\epsilon&\le 2\left(\frac{\pi(d-1)^2(3d-2)}{d\cdot n}\right)^2.
\end{align}
The probe state has dimension bounded as
\begin{align}\label{prob-dim}
d_{\rm P}\le \left(\frac{9n}{3d-2}\right)^{d^2-1}.
\end{align}
\end{theo}
Ref.\ \cite{kahn2007fast} showed that the estimation of an arbitrary $d$-dimensional unitary given $n$ uses can be done with an error scaling $ 1/n^2$. The error was measured by the entanglement gate infidelity, which is upper bounded by $1-(1-\epsilon)^2$. Theorem \ref{theo-opt-tomography} refines this result by not only achieving the $1/n^2$ scaling but also identifying an explicit expression of the constant of proportionality. In addition, our result holds for the more stringent error criterion $\epsilon$, i.e., the diamond norm error, and we also determine how the probe state dimension scales with $n$.

Combining Eq.\ (\ref{F-wc-tomo}) with Eq.\ (\ref{prob-dim}), we get:
\begin{cor}
The program cost $c_{\rm P}$ of Protocol \ref{alg-MO-program} is upper bounded as
\begin{align}
c_{\rm P}\le \left(\frac{d^2-1}{2}\right)\log\left(\frac{162\pi^2(d-1)^4}{d^2\cdot\epsilon}\right).
\end{align}
\end{cor}
It is obvious from the above corollary that 
\begin{align}\label{opt-scaling-dp-error}
c_{\rm P}\le \left(\frac{d^2-1}{2}\right)\log\left(\frac{162\pi^2d^2}{\epsilon}\right),
\end{align}
which matches Table \ref{table-compare} and achieves a quadratic reduction compared to known results.

\medskip

\noindent{\em Asymptotic equivalence of programming, metrology, and learning.}
From the previous discussion, we can see that an optimal way of programming a unitary is actually to let the processor learn and memorize it (see Figure~\ref{fig-mo}).
The task of learning a unitary $U$ from $n$ instances \cite{gammelmark2009quantum,bisio2010optimal,mo2019quantum} consists of a learning phase and an execution (or testing) phase.
In the learning phase, the protocol makes $n$ (not necessarily parallel) queries to $U$. In the execution phase, the protocol emulates the learned unitary on an arbitrary input state. Notice that the execution phase happens after the learning phase, thus the protocol should be able to store the information of $U$. 
%In Protocol \ref{alg-MO-program}, this is done by applying $U^{\otimes n}$ to the optimal state for metrology.

A learning protocol induces a programmable processor in the sense that the learning phase can be used to generate a program. Nevertheless, one should keep in mind  that learning and programming are not equivalent. 
Indeed, in the task of programming, the program does not have to be generated by learning, i.e., by applying multiple instances of $U$ on a quantum state.
%For example, the state $\theta/(2\pi) |0\>+\sqrt{1-\theta^2/(4\pi^2)}|1\>$ is a program for the phase gate $U_\theta:=|0\>\<0|+e^{-i\theta}|1\>\<1|$, yet it cannot be generated by applying multiple uses of $U_\theta$ to a quantum state. 
As learning has this additional constraint, its resource requirement is at least as stringent as that of programming. Therefore, since Protocol \ref{alg-MO-program} is an optimal processor, it is also an optimal learning protocol. The performance of optimal learning given $n$ instances is thus given by Theorem \ref{theo-converse}, achieved by unitary gate metrology. In summary, for finite dimensional quantum gates, the performances of programming, metrology and learning are asymptotically equal:
\begin{align*}
{\rm  programming \approx  metrology \approx learning. }
\end{align*}

\medskip
\noindent{\em Quantum versus classical advantage.} One may wonder if it is possible to simply use a classical program, e.g., to write down the description of the gate on a tape. Here we show, via a simple example, that our Protocol \ref{alg-MO-program}, which uses a quantum program, beats the best processor that uses classical programs in scaling. 

Let us consider the case of programming a phase gate $U_\theta=|0\>\<0|+e^{-i\theta}|1\>\<1|$, where $\theta\in[0,2\pi)$ is the (unknown) phase, for it allows for explicit calculations. Fixing the program dimension $d_{\rm P}:=2^{c_{\rm P}}$, the best classical strategy is nothing but dividing the range $[0,2\pi)$ into $d_{\rm P}$ equal-width intervals. The tag of the interval that contains $\theta$ is used as the program, and the processor runs $U_{\hat{\theta}}$ with $\hat{\theta}$ being the middle point of the interval. Since $\max|\hat{\theta}-\theta|=\pi/d_{\rm P}$, the error of this approach is $\epsilon_{\rm classical}=\sqrt{(1-\cos(\pi/d_{\rm P}))/2}\simeq\pi/(2d_{\rm P})$, which is inversely proportional to the program dimension.

In contrast, we can employ our Protocol \ref{alg-MO-program}, where we use the sine state \cite{buvzek1999optimal}
\begin{align}
|\psi\>=\sqrt{\frac{2}{d_{\rm P}}}\sum_{m=0}^{d_{\rm P}-1}\sin\frac{\pi(m+1/2)}{d_{\rm P}}|m\>.
\end{align}
as the probe state and the covariant POVM $\left\{\frac{\d\hat{\theta}}{2\pi}|\eta_{\hat{\theta}}\>\<\eta_{\hat{\theta}}|:|\eta_{\hat{\theta}}\>:=\sum_{m=0}^{d_{\rm P}-1}e^{-im\hat{\theta}}|m\>\right\}_{\hat{\theta}}$ as the measurement. The error can be evaluated as
\begin{align}
\epsilon_{\rm quantum}\simeq\frac{\pi^2}{2d_{\rm P}^2},
\end{align}
which is inversely proportional to the square of the program dimension. In other words, the program dimension of a processor with classical programs is quadratically larger than that of our quantum processor.  
In the more complex case of programming a $d$-dimensional unitary gate, the classical strategy is to construct an $\epsilon$-mesh of the unitary gates, which was employed by Ref.\ \cite{kubicki2019resource}. The program cost was given in Table~\ref{table-compare} as $d^2\log(K/\epsilon)$, higher than twice the cost of our quantum strategy in the small $\epsilon$ regime.
This proves the claimed quantum-over-classical advantage in programming.

\medskip
\noindent{\em Conclusion and further discussions.} We identified the optimal scaling of the program cost with accuracy in a universal quantum processor. The optimal scaling can be achieved with a measure-and-operate learning protocol. With this finding, we showed the asymptotic equivalence between programming, metrology, and learning.

In this work, we determined the optimal dependence of the program size on the accuracy parameter $\epsilon$. An interesting extension would be to determine the optimal scaling with the dimension of the target system $d$. Moreover, the task we focused on is universal programming, which requires the processor to work well for every gate of a certain dimension. 
%Would a smaller set of gates immediately lead to a smaller required program cost?
It is natural to expect that a smaller set of gates would lead to a smaller program cost. 
Observe from Eq.\ (\ref{opt-scaling-dp-error}) that the prefactor $(d^2-1)/2$ is exactly one half  the number of real parameters determining a qudit unitary gate (up to a global phase).
We therefore conjecture a general formula, valid for parametric families of quantum gates with a continuous dependence on $\nu$ real parameters:
\begin{align}
c_{\rm P}\sim \left(\frac{\nu}{2}\right)\log\left(\frac{C_{\nu,d}}{\epsilon}\right),
\end{align}
where $C_{\nu,d}$ is a parameter, possibly dependent on $\nu$ and $d$ but independent of $\epsilon$.

Another key reason for making this conjecture is that the ultimate performances of quantum information processing tasks share similar forms in the asymptotic limit of ``many copies''. In particular, one can consider the compression of identically prepared quantum systems, e.g.\ states of the form $\rho^{\otimes n}$ with $\rho$ unknown and $n$ being large. It turns out that the minimum cost of the memory, when requiring the error to be vanishing for large $n$, is $(\nu/2)\log n$ (qu)bits in the leading order \cite{plesch2010efficient,rozema2014quantum,chiribella2015universal,yang2016efficient,yang2016optimal,yang2018compression}. Here $\nu$, the number of variable real parameters, appears again.
Further pursuit in this direction could lead to the discovery of a universality rule, which governs the behaviour of optimal quantum devices  in the limit of macroscopically many copies.

\medskip

\begin{acknowledgments} 
We thank an anonymous reviewer of the conference ``QIP2021'' for a comment that  improves the scaling of the upper bound with respect to $d$. 
This work is supported by  the National Natural Science Foundation of China through grant 11675136, the Hong Kong Research Grant Council through grant 17300317,  the Foundational Questions Institute through grant FQXi-RFP3-1325, the Croucher Foundation,
 the AFOSR via grant No. FA9550-19-1-0202,
 the Swiss National Science Foundation via the National Center for Competence in Research ``QSIT" as well as via project No.\ 200020\_165843, and the ETH Pauli Center for Theoretical Studies. 

\end{acknowledgments} 

\bibliography{ref}

\appendix
\begin{widetext}

\section{Proof of Theorem 1}\label{app-thm-converse}

Consider any $\epsilon$-universal processor $(\map{C},\{\psi_{{\rm P},U}\})$. We prove Theorem 1 of the main text, which is a lower bound on the dimension $d_{\rm P}$ of the program, i.e.\ the dimension of $\Supp\{\psi_{{\rm P},U}\}$.

We first show that programming one use of $U$ with error $\epsilon$ requires the same amount of information as programming $m$ uses of $U$ with error $4m\sqrt{2\epsilon}$ for any $m\ge1$.
Note that the proof here extends that of \cite[Corollary 2]{chiribella2019energy}.
First, we define the \emph{worst-case input (or minimum) fidelity} between two arbitrary quantum channels $\map{A}$ and $\map{B}$, defined as \cite{nielsen2000quantum}
\begin{align}\label{F-wc}
F_{\rm wc}(\map{A},\map{B}):=\inf_{\Psi}F\left((\map{A}\otimes\map{I}_{\rm R})(\Psi),(\map{B}\otimes\map{I}_{\rm R})(\Psi)\right),
\end{align}
where the infimum is taken over all pure states $|\Psi\>\in\spc{H}_{\rm S}\otimes\spc{H}_{\rm R}$ with $\spc{H}_{\rm R}\simeq\spc{H}_{\rm S}$ being a reference system, and
$F(\rho,\sigma):=\left(\Tr\sqrt{\rho^{\frac12}\sigma\rho^{\frac12}}\right)^2$ is the Uhlmann fidelity for states. 
By this definition and the Fuchs - Van de Graaf inequality \cite{fuchs1999cryptographic}, we have 
\begin{align}
F_{\rm wc}\left(\map{E}_U,\map{U}\right)\ge (1-\epsilon)^2\ge 1-2\epsilon\qquad\forall\,U.
\end{align}

Denote by $\map{V}:\spc{H}\otimes\spc{H}_{\rm P}\to\spc{H}\otimes\spc{H}_{\rm P'}$ a Stinespring dilation of $\map{C}$, where $\spc{H}_{\rm P'}$ is an ancillary space. There exists a state $|\phi_{{\rm P'},U}\>\in\spc{H}_{\rm P'}$  such that 
\begin{align}
F_{\rm wc}\left(\map{V}\circ(\map{I}_{\rm S}\otimes\psi_{{\rm P},U}),\map{U}\otimes\phi_{{\rm P'},U}\right)\ge 1-2\epsilon.
\end{align}
Applying again the Fuchs - Van de Graaf inequality, we get
\begin{align}\label{app-gatedist1}
\left\|\map{V}\circ(\map{I}_{\rm S}\otimes\psi_{{\rm P},U})-\map{U}\otimes\phi_{{\rm P'},U}\right\|_{\diamond}\le 2\sqrt{2\epsilon}.
\end{align}
Notice that here $\psi_{{\rm P},U}$ is regarded as a channel that has trivial input and prepares the state $\psi_{{\rm P},U}$.

Next, define the pseudoinverse of $\map{V}$, $\map{E}_{{\rm inv},V}:\spc{H}_{\rm S}\otimes\spc{H}_{{\rm P'}}\to\spc{H}_{\rm S}\otimes\spc{H}_{\rm P}$, as the following quantum channel:
\begin{align}
\map{E}_{{\rm inv},V}(\cdot):=\map{V}^\dag\circ\map{P}_{\spc{H}_V}(\cdot)+\Tr\left[\map{P}_{\spc{H}^\perp_V}(\cdot)\right]\pi_{\spc{H}_{\rm S}\otimes\spc{H}_{\rm P}},
\end{align}
where for any  Hilbert space $\spc{K}$ we denote by $\pi_{\spc{K}}$ the maximally mixed state, $\spc{H}_V$ is the image of $V$, and $\map{P}_{\spc{H}_V}$ or $\map{P}_{\spc{H}_V^\perp}$ is the projection operation into $\spc{H}_V$ or $(\spc{H}_{\rm S}\otimes\spc{H}_{{\rm P'}})\setminus\spc{H}_V$. 
Then we have
\begin{align}
\map{V}\circ\map{E}_{{\rm inv},V}(\cdot)=\map{P}_{\spc{H}_V}(\cdot)+\Tr\left[\map{P}_{\spc{H}^\perp_V}(\cdot)\right]\pi_{\spc{H}_V}\ge\map{P}_{\spc{H}_V}(\cdot).
\end{align}
It follows that
\begin{align}
F_{\rm wc}\left(\map{V}\circ(\map{U}^\dag\otimes\psi_{{\rm P},U}),\map{P}_{\spc{H}_V}\circ(\map{I}_{\rm S}\otimes\phi_{{\rm P'},U})\right)\le F_{\rm wc}\left(\map{V}\circ(\map{U}^\dag\otimes\psi_{{\rm P},U}),\map{V}\circ\map{E}_{{\rm inv},V}\circ(\map{I}_{\rm S}\otimes\phi_{{\rm P'},U})\right).
\end{align}
Since $(V\otimes I_{\rm R})((U^\dag\otimes I_{\rm R})|\Psi\>\otimes|\psi_{{\rm P},U}\>)\in\spc{H}_V\otimes\spc{H}_{{\rm R}}$ for any input state $|\Psi\>\in\spc{H}\otimes\spc{H}_{\rm R}$, we have
\begin{align}
F_{\rm wc}\left(\map{V}\circ(\map{I}_{\rm S}\otimes\psi_{{\rm P},U}),\map{U}\otimes\phi_{{\rm P'},U}\right)&=F_{\rm wc}\left(\map{V}\circ(\map{U}^\dag\otimes\psi_{{\rm P},U}),\map{P}_{\spc{H}_V}\circ(\map{I}_{\rm S}\otimes\phi_{{\rm P'},U})\right)\\
&\le F_{\rm wc}\left(\map{V}\circ(\map{U}^\dag\otimes\psi_{{\rm P},U}),\map{V}\circ\map{E}_{{\rm inv},V}\circ(\map{I}_{\rm S}\otimes\phi_{{\rm P'},U})\right)\\
&= F_{\rm wc}\left(\map{U}^\dag\otimes\psi_{{\rm P},U},\map{E}_{{\rm inv},V}\circ(\map{I}_{\rm S}\otimes\phi_{{\rm P'},U})\right),
\end{align}
having used the property that $\map{V}$ (as an isometry) preserves fidelity in the last step.
Applying again the Fuchs - Van de Graaf inequality, we get 
\begin{align}\label{app-gatedist2}
\left\|\map{E}_{{\rm inv},V}\circ(\map{I}_{\rm S}\otimes\phi_{{\rm P'},U})-\map{U}^\dag\otimes\psi_{{\rm P},U}\right\|_{\diamond}\le 2\sqrt{2\epsilon}.
\end{align}

Now, we apply $\map{V}$ and $\map{E}_{{\rm inv},V}$ separately on two replicas of the system.
Using Eqs.\ (\ref{app-gatedist1}) and (\ref{app-gatedist2}) as well as basic properties (the triangle inequality and the data processing inequality) of the diamond norm, we have
\begin{align}
\left\|\map{E}^{{\rm P'S_2}}_{{\rm inv},V}\circ\map{V}^{{\rm S_1P}}\circ(\map{I}_{\rm S_1}\otimes\psi_{{\rm P},U}\otimes\map{I}_{\rm S_2})-\map{U}\otimes\psi_{{\rm P},U}\otimes\map{U}^\dag\right\|_{\diamond}\le 4\sqrt{2\epsilon},
\end{align}
where the superscript in $\map{V}^{{\rm S_1P}}$ indicates the registers that $\map{V}$ acts upon. 
Repeating this procedure for $m$ times and discarding the program in the end, we get a cascade of channels which acts on $2m$ replicas $\rm S_1,S_2,\dots,S_{2m}$ of the system:
\begin{align}
\tilde{\map{M}}_U(\cdot):=\Tr_{\rm P}\circ\,\map{E}^{{\rm P'S_{2m}}}_{{\rm inv},V}\circ\map{V}^{{\rm S_{2m-1}P}}\circ\cdots\circ\map{E}^{{\rm P'S_2}}_{{\rm inv},V}\circ\map{V}^{{\rm S_1P}}\left((\cdot)\otimes\psi_{{\rm P},U}\right)
\end{align}
whose distance from $m$ uses of the unitary channel $\map{U}\otimes\map{U}^\dag$ is bounded as
\begin{align}
\left\|\tilde{\map{M}}_U-\left(\map{U}\otimes\map{U}^\dag\right)^{\otimes m}\right\|_{\diamond}\le 4m\sqrt{2\epsilon}.
\end{align}
For simplicity of calculation, we now discard half of the systems $\{{\rm S}_{2j}\}_{j=1}^m$ in the above formula, obtaining
\begin{align}\label{app-approximate-mgates}
\left\|\map{M}_U-\map{U}^{\otimes m}\right\|_{\diamond}\le 4m\sqrt{2\epsilon}
\end{align}
with
\begin{align}
\map{M}_U(\cdot):=\Tr_{{\rm P,S_2,\dots,S_{2m}}}\circ\,\map{E}^{{\rm P'S_{2m}}}_{{\rm inv},V}\circ\map{V}^{{\rm S_{2m-1}P}}\circ\cdots\circ\map{E}^{{\rm P'S_2}}_{{\rm inv},V}\circ\map{V}^{{\rm S_1P}}\left((\cdot)\otimes\psi_{{\rm P},U}\right).
\end{align}
This concludes the first part of the proof. Observe that, on one hand, all information in $\map{M}_U$ on $U$ comes from the program state; on the other hand, $\map{M}_U$ is $(4m\sqrt{2\epsilon})$-close to $m$ uses of $U$. By comparing the amount of information, we argue that the program state has to contain almost the same amount of information as $m$ uses of $U$, for any $m\ll 1/\sqrt{\epsilon}$.

Next, we make the above argument quantitative.
As a measure of information, we consider the Holevo information $\chi$ \cite{holevo1973bounds}, defined for an ensemble of quantum states $\{\rho_x,\d x\}_{x\in\set{X}}$ as 
\begin{align}
I_{\rm H}\left(\{\rho_x,\d x\}\right):=H\left(\int_{x\in\set{X}}\d x\,\rho_{x}\right)-\int_{x\in\set{X}}\d x\,H(\rho_x)
\end{align}
where $H$ denotes the von Neumann entropy. 
%It can be rewritten as the mutual information of a bipartite state $\rho_{XS}:=\int\d x |x\>\<x|\otimes\rho_{x}$, where $X$ is a classical register containing the value of $x$ and $S$ is the state register. From the monotonicity of the mutual information under data processing on $S$, we see that the Holevo information is also monotonically decreasing under data processing.

Now let us derive an upper bound of the Holevo information of the program.
Consider inputting an arbitrary state $\Phi_m$ to $\map{M}_U$. Notice that $\chi$ is non-increasing under data processing on the system side. We get
\begin{align}
I_{\rm H}\left(\{\psi_{{\rm P},U},\d U\}\right)=I_{\rm H}\left(\{\psi_{{\rm P},U}\otimes\Phi_m,\d U\}\right)\ge I_{\rm H}\left(\left\{\map{M}_U(\Phi_m),\d U\right\}\right),
\end{align}
where $\d U$ is the Haar measure of $\grp{SU(d)}$.

We choose $\Phi_m$ to maximise $I_{\rm H}\left(\left\{\map{U}^{\otimes m}(\Phi_m),\d U\right\}\right)$.  
By the Schur-Weyl duality (see, e.g., Ref.\ \cite{fulton2013representation}), the $m$-qudit Hilbert space can be decomposed as
\begin{align}\label{schur-weyl}
\spc{H}^{\otimes m}\simeq\bigoplus_{\lambda\in\grp{S}_m}\spc{H}_{\lambda}\otimes\spc{M}_{\lambda,m},
\end{align}
where $\grp{S}_m:=\{\lambda\in\N^{\times d}~|~|\lambda|:=\sum_i\lambda_i=m,\lambda_i\ge\lambda_j\,\forall\,i<j\}$, each $\lambda$ is called a \emph{Young diagram}, each $\spc{H}_\lambda$ is an irreducible subspace of $\grp{SU(d)}$ characterized by the Young diagram $\lambda$, and $\spc{M}_{\lambda,m}$ is the corresponding multiplicity subspace. With this decomposition, $m$ parallel uses of $U\in\grp{SU(d)}$ can also be decomposed as
\begin{align}\label{unitary-schur-weyl}
U^{\otimes m}\simeq\bigoplus_{\lambda\in\grp{S}_m}U_{\lambda}\otimes I_{c_{\lambda,m}},
\end{align} 
where $U_\lambda$ is the irreducible representation of $\grp{SU(d)}$ characterized by the Young diagram $\lambda$ and $I_{c_{\lambda,m}}$ is the identity of the corresponding multiplicity subspace.

To this end, we can define $\map{T}:\spc{H}^{\otimes m}\to\bigoplus_{\lambda}\spc{H}_{\lambda}$ to be the quantum channel that first incorporates the isometry $\spc{H}^{\otimes m}\to\bigoplus_{\lambda}\spc{H}_{\lambda}\otimes\spc{M}_{\lambda,m}$ and then discards the multiplicity parts $\{\map{M}_{\lambda,m}\}$. Since $\map{U}^{\otimes m}$ is invariant on the multiplicity subspace, we have 
\begin{align}
I_{\rm H}\left(\left\{\map{T}\circ\map{U}^{\otimes m}(\Phi_m),\d U\right\}\right)=I_{\rm H}\left(\left\{\map{U}^{\otimes m}(\Phi_m),\d U\right\}\right)
\end{align}
for any $\Phi_m$.
The point of applying $\map{T}$ is that the dimension is reduced from $d^{2m}$ to 
\begin{align}
d_m&:=\sum_{\lambda\in\grp{S_m}}d_\lambda^2\nonumber\\
&={m+d^2-1\choose d^2-1}\label{Dn}
\end{align}
with $d_\lambda$ being the dimension of $\spc{H}_\lambda$, having used \cite[Eq.~(57)]{schur1901klasse} in the second equality. It is obvious that $d_m$ grows only polynomially instead of exponentially in $m$. Explicitly, we have
\begin{align}\label{dm-bnd}
d_m\ge\left(\frac{m}{d^2-1}\right)^{d^2-1}.
\end{align}

We then take $\Phi_m$ to be
\begin{align}
|\Phi_m\>:=\bigoplus_{\lambda\in\grp{S}_m}\frac{d_\lambda}{\sqrt{d_m}}\left(|\Phi^+_{\lambda}\>\otimes|\psi_0\>\right)
\end{align}
where $|\Phi^+_\lambda\>\in\spc{H}_\lambda\otimes\spc{H}_{\lambda}$ is the maximally entangled state and $|\psi_0\>$ is an arbitrary state in the multiplicity spaces.
This choice of $\Phi_m$ achieves the maximum Holevo information
\begin{align}
I_{\rm H}\left(\left\{\map{T}\circ\map{U}^{\otimes m}(\Phi_{m}),\d U\right\}\right)=\log d_m.
\end{align}

Therefore, we have
\begin{align}
I_{\rm H}\left(\{\psi_{{\rm P},U},\d U\}\right)&\ge I_{\rm H}\left(\left\{\map{U}^{\otimes m}(\Phi_m),\d U\right\}\right)\\
&\ge I_{\rm H}\left(\left\{\map{U}^{\otimes m}(\Phi_m),\d U\right\}\right)-(4m\sqrt{2\epsilon})\log d_m-1\\
&=I_{\rm H}\left(\left\{\map{T}\circ\map{U}^{\otimes m}(\Phi_m),\d U\right\}\right)-(4m\sqrt{2\epsilon})\log d_m-1, \label{bound-chi-inter1}
\end{align} 
having used Eq.\ (\ref{app-approximate-mgates}) and the Fannes-Alicki-Winter inequality \cite{alicki2004continuity,winter2016tight} to get the second inequality.
Taking into account the bound $\log d_{\rm P}\ge I_{\rm H}\left(\{\psi_{{\rm P},U},\d U\}\right)$, the inequality (\ref{bound-chi-inter1}) becomes
\begin{align}\label{app-bound-chi-inter2}
\log d_{\rm P}&\ge(1-4m\sqrt{2\epsilon})\log d_m-1.
\end{align}
For an arbitrarily $\epsilon$-independent parameter $\delta>0$, we choose 
\begin{align}
m=\left\lceil\frac{\delta}{4\sqrt{2\epsilon}}\right\rceil,
\end{align}
where $\lceil\cdot\rceil$ denotes the ceiling function.
Substituting this choice of $m$ as well as Eq.~(\ref{dm-bnd}) into the bound, we get
\begin{align}
d_{\rm P}&\ge \frac{1}{2}\left(\frac{\delta}{4\sqrt{2\epsilon}(d^2-1)}\right)^{(1-\delta-4\sqrt{2\epsilon})(d^2-1)}.\label{program-dim-bound-explicit}
\end{align}
With this, we conclude that, for any $\alpha<(d^2-1)/2$, we have
\begin{align}
d_{\rm P}&=\Omega\left(1/\epsilon^\alpha\right).
\end{align}
\qed
\iffalse

The bound in Eq.\ (\ref{program-dim-bound-explicit}) decreases with $d$, which seems to contradict the intuition that it is more demanding to program a larger unitary gate.
In fact, we can show an alternative bound that increases with $d$ by modifying a few steps in the above proof. 

First, observe that Lemma \ref{app-ansatz-Dscaling} does not work very well in the case of fixed $m$ and large $d$. Instead of using Lemma \ref{app-ansatz-Dscaling} in Eq.\ (\ref{app-bound-chi-inter2}), we employ directly the definition (\ref{Dn}):
\begin{align}
\log d_{\rm P}&\ge (1-4m\sqrt{2\epsilon})\log\left(\sum_{\lambda\in\grp{S_m}}d_\lambda^2\right)-1\\
&\ge(1-4m\sqrt{2\epsilon})\log\left(d_{(m,0,\dots,0)}\right)^2-1\\
&=2(1-4m\sqrt{2\epsilon})\log{d+m-1\choose m-1}-1.
\end{align}
Suppose $\epsilon$ is small enough, e.g., $\epsilon=0.001$. We take $m$ to be the closest integer to $1/(8\sqrt{2\epsilon})$ in the above inequality, which becomes
\begin{align}
\log d_{\rm P}&\ge \log{d+1/(8\sqrt{2\epsilon})-1\choose 1/(8\sqrt{2\epsilon})-1}-1.
\end{align}
Therefore, we get an alternative bound
\begin{align}
 d_{\rm P}& \ge \frac12{d+1/(8\sqrt{2\epsilon})-1\choose 1/(8\sqrt{2\epsilon})-1}\sim d^{\frac{1}{8\sqrt{2\epsilon}}-1}.
\end{align} 
 \fi
\section{Proof of Theorem 2}\label{app-thm-tomography}

In this section we prove Theorem 2 of the main text on the performance of qudit gate estimation. 
The estimation task consists of two steps: The first step is to prepare a suitable \emph{probe state} $|\psi\>$ and then to apply $n$ parallel uses of $U$ on it. The second step is to measure the resultant state, denoted by $|\psi_{U,n}\>$, with a suitable POVM $\{M_{\hat{U}}\}_{\hat{U}\in\grp{SU(d)}}$, which outputs an estimate $\hat{U}$ of $U$. 

Here we measure the performance of unitary gate metrology by the diamond norm error:
\begin{align}
\epsilon:=\sup_{U\in\grp{SU(d)}}\frac12\left\|\map{U}-\map{E}_{{\rm mo},U}\right\|_{\diamond}.
\end{align}
Here $\map{E}_{{\rm mo},U}$ is the measure-and-operate (MO) channel
\begin{align}
\map{E}_{{\rm mo},U}(\cdot):=\int\d\hat{U}\,p(\hat{U}|U)\,\hat{\map{U}}(\cdot),
\end{align}
where $p(\hat{U}|U)$ is the probability of getting the estimate $\hat{U}$ (when the actual gate is $U$) defined as
\begin{align}\label{est-prob-dist}
p(\hat{U}|U):=\Tr\left[\psi_{U,n}M_{\hat{U}}\right].
\end{align}
We remark that the performance of unitary gate metrology can also be characterised by other figures of merit, e.g., the (average) gate fidelity \cite{horodecki1999general,nielsen2002simple}. Here we are using a more demanding error measure.

The proof can be sketched as the following:
\begin{enumerate}
\item We first measure the performance of estimation protocols using the \emph{entanglement fidelity} \cite{raginsky2001fidelity}:
\begin{align}\label{F-ent}
F_{\rm ent}(\map{A},\map{B}):=F\left((\map{A}\otimes\map{I}_{\rm R})(|\Phi^+\>\<\Phi^+|),(\map{B}\otimes\map{I}_{\rm R})(|\Phi^+\>\<\Phi^+|)\right),
\end{align}
where $|\Phi^+\>$ is the maximally entangled state of the system ${\rm S}$ and a reference ${\rm R}\simeq{\rm S}$.  In general, $F_{\rm ent}$ serves as an upper bound on $F_{\rm wc}$ and is easier to evaluate.
\item We derive a formula of $F_{\rm ent}$ for a class of estimation protocols, which include the optimal protocol that achieves the maximum of $F_{\rm ent}$ over all protocols. The optimal protocol and its $F_{\rm ent}$ can be evaluated numerically from the formula.
\item Next, we show that, for the above class of protocols, $\epsilon =1-F_{\rm ent}$.
\item We fix an estimation protocol and prove that it achieves the performance $F_{\rm ent}\ge 1-c_d/n^2$. Combining with the point above, we obtain an upper bound on $\epsilon$ in terms of $n$.
\item We also determine, for the same protocol, the relation between the dimension of the probe and $n$.
\end{enumerate}

\subsection{A formula for $F_{\rm ent}$}\label{app-subsec-optimal-tomo}

In this subsection, we focus first on the entanglement fidelity $F_{\rm ent}$. Before starting, we recall a few concepts from the Schur-Weyl decomposition [cf.\ Eq.\ (\ref{schur-weyl})].
We will make frequent uses of the Young diagrams $\lambda=(\lambda_1,\lambda_2,\dots)$ and the irreducible representation $U_{\lambda}$ characterised by the Young diagram $\lambda$ [see Eq.\ (\ref{unitary-schur-weyl})]. In particular, we define $e_i$ to be the vector whose $i$-th entry is one and other entries are zero. By definition, $e_1$ corresponds to a legitimate Young diagram whose associated representation is the $d$-dimensional self-representation, and we use the abbreviation $U:=U_{e_1}$. 
We will use the double-ket notation $|A\kk:=\sum_{n,m}\<n|A|m\>|n\>\<m|$ ($\{|n\>\}$ being an orthonormal basis) for a matrix $A$ and denote by $|\Phi^+_{U,\lambda}\>$ the maximally entangled state $|U_{\lambda}\kk/\sqrt{d_\lambda}$.

To maximise the entanglement fidelity of metrology, it is enough to consider probe states of the form \cite[Theorem 1]{chiribella2005optimal}
\begin{align}\label{input-state-form}
|\psi\>=\bigoplus_{\lambda\in\set{S}_{\rm Young}}\sqrt{q_\lambda}|\Phi^+_{\lambda}\>\otimes|\Phi^+_{m_\lambda}\>.
\end{align}
Here $\set{S}_{\rm Young}\subset\grp{S_n}$ is a suitable set containing Young diagrams of $n$ boxes, $|\Phi^+\>$ denotes the maximally entangled state (of the corresponding Hilbert spaces), and $\{q_\lambda\}$ is a suitable probability distribution. We assume that any Young diagram $\lambda\in\set{S}_{\rm Young}$ has strictly decreasing row numbers.
After the application of $U^{\otimes n}$, the probe state is in the form
\begin{align}
|\psi_{U,n}\>=\bigoplus_{\lambda\in\set{S}_{\rm Young}}\sqrt{q_\lambda}|\Phi^+_{U,\lambda}\>\otimes|\Phi^+_{m_\lambda}\>.
\end{align}
The optimal measurement \cite{chiribella2005optimal} is the covariant POVM $\{\d\hat{U}, M_{\hat{U}}\}$ with $\d\hat{U}$ being the Haar measure and 
\begin{align}\label{optimal-POVM}
M_{\hat{U}}:=|\eta_{\hat{U}}\>\<\eta_{\hat{U}}|\qquad|\eta_{\hat{U}}\>:=\bigoplus_{\lambda\in\set{S}_{\rm Young}}d_{\lambda}|\Phi^+_{\hat{U},\lambda}\>\otimes |\Phi^+_{m_\lambda}\>.
\end{align}
Denoting by $\chi_{U,\lambda}:=\Tr[U_{\lambda}]$ the characters of $\grp{SU(d)}$, the probability of getting the outcome $\hat{U}$ when the actual gate is $U$ can be expressed as
\begin{align}\label{est-prob-dist-covariant}
p(\hat{U}|U)=\left|\sum_{\lambda\in\set{S}_{\rm Young}}\sqrt{q_\lambda}\chi_{U\hat{U}^{-1},\lambda}\right|^2.
\end{align}

We can then express the entanglement fidelity as
\begin{align}
F_{\rm ent}\left(\map{E}_{{\rm mo},U},\map{U}\right)=\inf_{U\in\grp{SU(d)}}\frac{1}{d^2}\int \d \hat{U} \left|\chi_{U\hat{U}^{-1}}\sum_{\lambda\in\set{S}_{\rm Young}}\sqrt{q_\lambda}\chi_{U\hat{U}^{-1},\lambda}\right|^2.
\end{align}
where $\chi_{U\hat{U}^{-1}}(:=\chi_{U\hat{U}^{-1},e_1})$ is the character of the self-representation $e_1$.
To proceed, we decompose the characters as
\begin{align}
\chi_{U\hat{U}^{-1}}\chi_{U\hat{U}^{-1},\lambda}=\sum_{\lambda'\in\set{O}_1(\lambda)}\chi_{U\hat{U}^{-1},\lambda'},
\end{align}
where $\set{O}_1(\lambda):=\left\{\lambda+e_i~|~  i:\lambda_i<\lambda_{i-1}\right\}$. 
Using the group invariance property of the Haar measure and the orthogonality of characters, we have
\begin{align}
F_{\rm ent}\left(\map{E}_{{\rm mo},U},\map{U}\right)&=\inf_{U}\frac{1}{d^2}\int \d \hat{U} \left|\sum_{\lambda\in\set{S}_{\rm Young}}\sqrt{q_\lambda}\sum_{\lambda'\in\set{O}_1(\lambda)}\chi_{U\hat{U}^{-1},\lambda'}\right|^2\\
&=\inf_{U}\frac{1}{d^2} \left(\sum_{\lambda,\tilde{\lambda}\in\set{S}_{\rm Young}}\sqrt{q_\lambda q_{\tilde{\lambda}}}\sum_{\lambda'\in\set{O}_1(\lambda),\tilde{\lambda}'\in\set{O}_1(\tilde{\lambda})}\int \d \hat{U}~\chi_{U\hat{U}^{-1},\lambda'}\chi_{U\hat{U}^{-1},\tilde{\lambda}'}^*\right)\\
&=\frac{1}{d^2} \left(\sum_{\lambda,\tilde{\lambda}\in\set{S}_{\rm Young}}\sqrt{q_\lambda q_{\tilde{\lambda}}}\sum_{\lambda'\in\set{O}_1(\lambda),\tilde{\lambda}'\in\set{O}_1(\tilde{\lambda})}\delta_{\lambda'\tilde{\lambda}'}\right).
\end{align}
Rearranging terms, we have
\begin{align}
F_{\rm ent}\left(\map{E}_{{\rm mo},U},\map{U}\right)&=\frac{1}{d^2}\sum_{\lambda'\in\grp{S_{n+1}}}\left(\sum_{\lambda\in\set{O}_{\lambda'}}\sqrt{q_\lambda}\right)^2,
\end{align}
where $\set{O}_{\lambda'}:=\left\{\lambda\in\set{S}_{\rm Young}~|~\exists\,i,\ \lambda'=\lambda+e_i\right\}$.
Equivalently, the entanglement fidelity can be expressed as 
\begin{align}\label{optimal-estimation-fidelity}
F_{\rm ent}\left(\map{E}_U,U\right)=\frac{1}{d^2}\left(\vec{q}^T S\vec{q}\,\right),
\end{align}
where $\vec{q}$ is a unit vector (i.e.\,$\vec{q}\cdot\vec{q}=1$) supported by $\set{S}_{\rm Young}$ and $S$ is the score matrix defined by
\begin{align}\label{def-score-matrix}
S_{\lambda\lambda'}:=\left\{\begin{array}{cc}d \qquad &  d_{\rm Young}(\lambda,\lambda')=0\\ & \\ 1 \qquad &  d_{\rm Young}(\lambda,\lambda')=2\\ & \\ 0\qquad &{\rm else}\end{array}\right..
\end{align}  
Here $d_{\rm Young}(\lambda,\lambda'):=\sum_i |\lambda_i-\lambda'_i|$ is a distance measure between Young diagrams.
Summarizing the above derivation, we have shown that:
\begin{lemma}\label{lemma-opt-tomo-fid}
Assume that any Young diagram $\lambda\in\set{S}_{\rm Young}$ has strictly decreasing row numbers. The entanglement fidelity of the optimal estimation is given by the optimization in Eq.\ (\ref{optimal-estimation-fidelity}).
\end{lemma}
The same result, in a slightly different form, was first obtained by Kahn \cite{kahn2007fast}. We remark that, though the optimal estimation performance is just the maximum eigenvalue of $S$ (\ref{def-score-matrix}), it is not easy to show the $1/n^2$ error scaling. The matrix $S$ is a banded multilevel Toeplitz matrix, whose eigensystem problem remains open to the best of our knowledge (see, e.g., Ref.\ \cite{ekstrom2018eigenvalues}).  

\subsection{Switching between the diamond norm error $\epsilon$ and $F_{\rm ent}$ for covariant protocols}
Here we show that for any covariant estimation protocol, defined as follows, it is enough to evaluate the entanglement fidelity: 
\begin{defi}[Covariant estimation protocols]
An estimation protocol $(\psi,\{M_{\hat{U}}\})$ is covariant if the probability distribution (\ref{est-prob-dist}) of the estimate satisfies
\begin{align}\label{prob-covariant}
p(W\hat{U}V^\dag|WUV^\dag)=p(\hat{U}|U)\qquad\forall\,W,V\in\grp{SU(d)}.
\end{align}
\end{defi}
One can directly check that protocols mentioned in the previous subsection, whose $p(\hat{U}|U)$ has the form (\ref{est-prob-dist-covariant}), are covariant. For covariant protocols, the channel $\map{E}_{{\rm mo},U}$ is covariant when $U=I$, and we have the following lemma:

\begin{lemma}\label{lemma-wc-equal-ent}
For any covariant estimation protocol, the following bound holds
\begin{align}
\epsilon= 1-F_{\rm ent}\left(\map{E}_{{\rm mo},I},\map{I}\right).
\end{align}
Therefore, it is enough to consider the quantity $F_{\rm ent}\left(\map{E}_{{\rm mo},I},\map{I}\right)$.
\end{lemma}
\medskip

\noindent{\bf Proof of Lemma \ref{lemma-wc-equal-ent}.} The proof consists of two steps: the first is to show that $\map{E}_{{\rm mo},I}$ is covariant (even though $\map{E}_{{\rm mo},U}$ is not in general), and the second is to make use of the symmetry induced by covariance to reduce the diamond norm error to $1-F_{\rm ent}$. Note that the techniques used in the second step has already been exploited in a couple of previous works \cite{matsumoto2012input,majenz2018entropy,pirandola2019fundamental,yang2020covariant}.

%First, from the expression of fidelity we have $F_{\rm wc}(\map{E}_{{\rm mo},U},\map{U})=F_{\rm wc}(\map{E}_{{\rm mo},U}\circ\map{U}^\dag)$ for any $U$.
Applying Eq.\ (\ref{prob-covariant}) we have
\begin{align}
\map{E}_{{\rm mo},U}\circ\map{U}^\dag&=\int\d\hat{U}\,p(\hat{U}_0U|U)\,\hat{\map{U}}_0\qquad \hat{U}_0:=\hat{U}U^\dag\\
&=\int\d\hat{U}\,p(\hat{U}_0|I)\,\hat{\map{U}}_0\\
&=\map{E}_{{\rm mo},I}.
\end{align}
Therefore, by unitary invariance of the diamond norm, we have $\epsilon=\frac12\|\map{E}_{{\rm mo},I}-\map{I}\|_{\diamond}$. What remains is to relate the diamond norm $\|\map{E}_{{\rm mo},I}-\map{I}\|_{\diamond}$ to the entanglement fidelity $F_{\rm ent}(\map{E}_{{\rm mo},I},\map{I})$.
Indeed, we have
\begin{align}
\map{E}_{{\rm mo},U}\circ\map{U}'&=\int \d\hat{U}\,p(\hat{U}|U)\,\hat{\map{U}}\circ\map{U}'\\
&=\int \d(U'\hat{U}_1U'^\dag)\, p\left(U'\hat{U}_1{U'}^\dag U^\dag|I\right)\,\map{U'}\circ\hat{\map{U}}_1\qquad\hat{U}_1:=U'^\dag\hat{U}U'\\
&=\map{U}'\circ\map{E}_{{\rm mo},U'^\dag UU'}.
\end{align}
for any $U'$. Taking $U$ to be the identity, it is immediate that $\map{E}_{{\rm mo},I}$ is covariant with respect to $\grp{SU(d)}$, i.e.\ $\map{E}_{{\rm mo},I}\circ\map{U}'=\map{U}'\circ\map{E}_{{\rm mo},I}$.
For covariant channels, we have the following general result:

%the worse-case input fidelity is equal to the entanglement fidelity \cite{matsumoto2012input}, i.e.\ 
%\begin{align}
%F_{\rm wc}\left(\map{E}_{{\rm mo},I},\map{I}\right)=F_{\rm ent}\left(\map{E}_{{\rm mo},I},\map{I}\right).
%\end{align}

For any quantum channel $\map{A}$ acting on a $d$-dimensional system, define its Choi state as
\begin{align}
	A := \Big(\map{A} \otimes \map{I} \Big) (\Phi^+) \, 
\end{align}
with $\Phi^+$ being the maximally entangled state in $\spc{H}\otimes\spc{H}$.
When $\map{A}$ is covariant, we have
\begin{align}
	[A, \, U \otimes U^\ast] = 0, \qquad \forall\, U \in {\rm SU}(d) \, .
\end{align}
By Schur's lemma, the Choi state of a covariant channel $\map{A}$ can be decomposed as 
\begin{align}
&A = (1-a) \cdot  \Phi^+ + a\cdot \rho^{\perp}\qquad\rho^{\perp}:=\frac{1}{d^2-1} \Big(I\otimes I - \Phi^+\Big)
\end{align}
for some $a\in [0,1]$. It follows immediately from the above expression that
\begin{align}
1-F_{\rm ent}(\map{A},\map{I})=\frac12\left\|A-\Phi^+\right\|_1,
\end{align}
which is the trace distance error between the Choi state of $\map{A}$ and the maximally entangled state. 
Finally, since for covariant channels the optimal input for the diamond norm is just the maximally entangled state \cite{matsumoto2012input}, i.e., $\|\map{A}-\map{I}\|_{\diamond}=\|A-\Phi^+\|_1$ we have the   equality as desired.

\iffalse
Combining with the inequality $\|\map{A}-\map{I}\|_{\diamond}\le d\cdot\|A-\Phi^+\|_1$ (see, e.g., \cite[Exercise 3.6]{watrous2018theory}), we get 
\begin{align}
\frac12\|\map{A}-\map{I}\|_{\diamond}\le d\cdot\left(1-F_{\rm ent}(\map{A},\map{I})\right)
\end{align}
as desired.
\fi
\qed

\subsection{Proof of Eq.\ (5) of the main text}
Now, we show that there exists a covariant protocol with worst-case fidelity given by Eq.\ (5) of the main text. Due to Lemma \ref{lemma-wc-equal-ent}, it is enough to show the bound for the entanglement fidelity.

The covariant estimation protocol we are going to discuss is of the structure described previously: Its input state is of the form (\ref{input-state-form}), its POVM is given by Eq.\ (\ref{optimal-POVM}), and its entanglement fidelity is given by Eq.\ (\ref{optimal-estimation-fidelity}). What remains to be done is to specify the distribution $\{q_\lambda\}$.

For this purpose, we first define a parameter $N$ that depends on $n$ as
\begin{align}\label{N}
N=\left\lfloor\frac{1}{(3d-2)}\left(\frac{2n}{d-1}+d-2\right)\right\rfloor
\end{align}
and $n_0:=n-((3d-2)N-d+2)(d-1)/2$. By definition, $N$ is bounded as
\begin{align}\label{N-bound}
N\in \left[c_{\min,d}\cdot n,c_{\max,d}\cdot n\right],\qquad c_{\min,d}:=\frac{2\left(1-\frac{d(d-1)}{n}\right)}{(3d-2)(d-1)}\quad c_{\max,d}:=\frac{2\left(1+\frac{(d-2)(d-1)}{2n}\right)}{(3d-2)(d-1)}
\end{align}
with $c_{\min,d}$ and $c_{\max,d}$ depending only on $d$ when $n\to\infty$.
Define $\mu_0\in\grp{S}_{n_0}$ as the most flat Young diagram with $n_0$ boxes:
\begin{align}\label{flat-Young}
\mu_0:=(\mu_{0,1},\mu_{0,2},\dots,\mu_{0,d})\quad{\rm s.t.}\ \sum_i|\mu_{0,i}|=n_0\quad{\rm and}\quad\mu_{0,j}+1\ge\mu_{0,i}\ge\mu_{0,j}\quad\forall\,j>i.
\end{align}
%{\color{red} Since it is a Young diagram, $\mu_{0,i}\ge\mu_{0,j}\quad\forall\,j>i$ should hold.}
Now we define the following viable subset of Young diagrams with $d$ rows and $n$ boxes, on which our probe state has support:
\begin{align}\label{viable-Young}
\set{S}_{\rm Young}:=\left\{\lambda\in\set{S}_n~|~\lambda_i=\mu_{i,0}+N(2d-3)+1-(N+1)(i-1)+\tilde{\lambda}_i,\,\forall i\le d-1\quad\exists\tilde{\lambda}\in[N-1]^{\times(d-1)}\right\}.
\end{align}
Obviously, the above definition satisfies the assumption that any Young diagram $\lambda\in\set{S}_{\rm Young}$ has strictly decreasing row numbers.
This choice is to minimise the boundary set, which contains those elements of $\set{S}_{\rm Young}$ with some of their adjacent (i.e. $d_{\rm Young}=2$) Young diagrams not in the set. One can see from Eq. (\ref{def-score-matrix}) that this makes the score higher. Moreover, as shown later, dimensions of elements in $\set{S}_{\rm Young}$ are easy to bound.
%Moreover, as shown later, all elements in $\set{S}_{\rm Young}$ has the same asymptotic dimension up to a constant, which makes the overall dimension easy to evaluate.

Each Young diagram in $\set{S}_{\rm Young}$ is now uniquely characterized by $\tilde{\lambda}\in[N-1]^{\times(d-1)}$, so from now on we use $\tilde{\lambda}$ as the notion for Young diagrams.
Note that the relevant elements of $S$ for the Young diagrams we consider are
\begin{align}
S(\tilde{\lambda},\tilde{\lambda}')=\left\{\begin{array}{ll} 
1 \qquad &  \tilde{\lambda}-\tilde{\lambda}'=\pm f_{ij}\quad\exists\,i>j\\ & \\
1 \qquad &  \tilde{\lambda}-\tilde{\lambda}'=\pm e_i\quad\exists\,i\\ & \\
d \qquad &  \tilde{\lambda}=\tilde{\lambda}'\\ & \\ 
0\qquad &{\rm else}\end{array}\right..
\end{align}
Here $\{e_i\}_{i=1}^{d-1}$ is the natural basis of $[N-1]^{\times d-1}$, and $f_{ij}:=e_i-e_j$.
We denote by $g_k$ the following distribution over $[N-1]$
\begin{align}\label{g}
g_{k}:=\frac{2}{N}\sin^2\left(\frac{\pi(2k+1)}{2N}\right)
\end{align}
and by $\epsilon_g$ the quantity 
\begin{align}\label{scaling-epsilon-g}
\epsilon_g:=1-\sum_{k=0}^{N-2}\sqrt{g_k g_{k+1}}\le\frac{\pi^2}{N^2}.
\end{align}
The inequality can be shown by straightforward calculation.
Consider the product form distribution
\begin{align}\label{def-q}
q^\ast_{\tilde{\lambda}}:=\prod_{i=1}^{d-1}g_{\tilde{\lambda}_i},
\end{align}
where $g$ is the distribution defined in Eq.\ (\ref{g}).

Now, we show that the covariant protocol, specified by Eq.\ (\ref{def-q}), has entanglement fidelity as follows. 
\begin{lemma}\label{lemma-opt-tomography}
The entanglement fidelity of the protocol specified by Eq.\ (\ref{def-q}) is lower bounded as
\begin{align}
F_{\rm ent}\left(\map{E}^\ast_{{\rm mo},I},\map{I}\right)\ge 1-2\left(\frac{\pi(d-1)}{d\cdot c_{\min,d}\cdot n}\right)^2,
\end{align}
where $c_{\min,d}$ is given in Eq.\ (\ref{N-bound}).
\end{lemma}

\noindent{\bf Proof of Lemma \ref{lemma-opt-tomography}.}
By definition (\ref{def-q}), we have
\begin{align}
\vec{q^\ast}^TS\vec{q^\ast}&=\sum_{\tilde{\lambda},\tilde{\lambda}'\in[N-1]^{\times (d-1)}}\sqrt{q^\ast_{\tilde{\lambda}}}S(\tilde{\lambda},\tilde{\lambda}')\sqrt{q^\ast_{\tilde{\lambda}'}}\\
&=\sum_{\tilde{\lambda}\in[N-1]^{\times (d-1)}}\sum_{i\not=j}\sqrt{q^\ast_{\tilde{\lambda}}q^\ast_{\tilde{\lambda}+f_{ij}}}+\sum_{\tilde{\lambda}\in[N-1]^{\times (d-1)}}\sum_{i}\sqrt{q^\ast_{\tilde{\lambda}}q^\ast_{\tilde{\lambda}+e_{i}}}+\sum_{\tilde{\lambda}\in[N-1]^{\times (d-1)}}\sum_{i}\sqrt{q^\ast_{\tilde{\lambda}}q^\ast_{\tilde{\lambda}-e_{i}}} +d\\
&=\sum_{i\not=j}\sum_{\tilde{\lambda}\in[N-1]^{\times (d-1)}}\sqrt{q^\ast_{\tilde{\lambda}}q^\ast_{\tilde{\lambda}+f_{ij}}}+\sum_{i}\left(\sum_{\tilde{\lambda}\in[N-1]^{\times (d-1)}}\sqrt{q^\ast_{\tilde{\lambda}}q^\ast_{\tilde{\lambda}+e_{i}}}+\sum_{\tilde{\lambda}\in[N-1]^{\times (d-1)}}\sqrt{q^\ast_{\tilde{\lambda}}q^\ast_{\tilde{\lambda}-e_{i}}}\right) +d.\label{fid-inter1}
\end{align}
For an arbitrary pair of $(i,j)$ such that $i\not =j$, using Eqs. (\ref{g}), (\ref{scaling-epsilon-g}) and (\ref{def-q}) we can explicitly evaluate the term in the first summation as
\begin{align}
\sum_{\tilde{\lambda}\in[N-1]^{\times (d-1)}}\sqrt{q^\ast_{\tilde{\lambda}}q^\ast_{\tilde{\lambda}+f_{ij}}}=\left(\sum_{\tilde{\lambda}_i=0}^{N-2}\sqrt{g_{\tilde{\lambda}_i}g_{\tilde{\lambda}_i+1}}\right)\left(\sum_{\tilde{\lambda}_j=1}^{N-1}\sqrt{g_{\tilde{\lambda}_j}g_{\tilde{\lambda}_j-1}}\right)=(1-\epsilon_g)^2.
\end{align}
Similarly, for arbitrary $i$, the term in the second and summation can be expressed as
\begin{align}
\sum_{\tilde{\lambda}\in[N-1]^{\times (d-1)}}\sqrt{q^\ast_{\tilde{\lambda}}q^\ast_{\tilde{\lambda}+e_{i}}}+\sum_{\tilde{\lambda}\in[N-1]^{\times (d-1)}}\sqrt{q^\ast_{\tilde{\lambda}}q^\ast_{\tilde{\lambda}-e_{i}}}=2(1-\epsilon_g).
\end{align}
Substituting the above back into Eq.\ (\ref{fid-inter1}), we have
\begin{align}
\vec{q^\ast}^TS\vec{q^\ast}&=d+(d-1)(d-2)(1-\epsilon_g)^2+2(d-1)(1-\epsilon_g)\\
&\ge d^2-2(d-1)^2\epsilon_g.
\end{align}
Combining the above inequality with Lemma \ref{lemma-opt-tomo-fid} and Eqs.\ (\ref{N-bound}) and (\ref{scaling-epsilon-g}), we get the bound
\begin{align}
F_{\rm ent}\left(\map{E}^\ast_{{\rm mo},I},\map{I}\right)&\ge 1-\frac{2\pi^2(d-1)^2}{d^2}\cdot\left(\frac{1}{N}\right)^2\\ 
&=1-2\left(\frac{\pi(d-1)}{d\cdot c_{\min,d}\cdot n}\right)^2
\end{align}
 as desired.\qed

\subsection{Proof of Eq.\ (6) of the main text}
We conclude our proof of Theorem 2 of the main text by showing Eq.\ (6) of the main text, which is a bound on the dimension of the probe state (\ref{def-q}).

\begin{lemma}\label{app-lemma-dim-probe}
The probe state specified by Eq.\ (\ref{def-q}) has dimension bounded as
\begin{align}
d_{\rm P}\le \left(2(d-1)c_{\max,d}\cdot n+3\right)^{d^2-1},
\end{align}
where $c_{\max,d}$ is given in Eq.\ (\ref{N-bound}).
\end{lemma}
\medskip

\noindent{\bf Proof of Theorem 2 of the main text.} Finally, putting together all ingredients (Lemmas \ref{lemma-wc-equal-ent}, \ref{lemma-opt-tomography}, and \ref{app-lemma-dim-probe}) yields Theorem 2 of the main text. We also used the bounds $c_{\min,d}\ge\frac{1}{(3d-2)(d-1)}$, $c_{\max,d}\le\frac{3}{(3d-2)(d-1)}$ and $3\le 3n/(3d-2)$, which come from the assumptions on $n$ and $d$, to simplify the expressions.
\qed

\medskip

\noindent{\bf Proof of Lemma \ref{app-lemma-dim-probe}.}
The irreducible representation $\lambda$ of $\grp{SU(d)}$ has dimension \cite[Eq.~(III.10)]{itzykson1966unitary}
\begin{align}\label{d-lambda}
d_{\lambda}=\frac{\prod_{1\le i<j\le d}(\lambda_i-\lambda_j-i+j)}{\prod_{k=1}^{d-1}k!}.
\end{align}
The viable set $\set{S}_{\rm Young}$ [cf.\ Eq.\ (\ref{viable-Young})] is so defined that, for any $\lambda\in\set{S}_{\rm Young}$ and any $i<j$,
\begin{align}
\lambda_i-\lambda_j\le\left\{\begin{matrix} N(j-i+1)+(2j-2i-1) &\qquad j<d\\ N(2j-i-1)+(j-2i+1)& \qquad j=d\end{matrix}\right.
\end{align}
Therefore, using Eq.\ (\ref{N-bound}), for any $\tilde{\lambda}\in\set{S}_{\rm Young}$, its dimension is upper bounded by
\begin{align}
d_{\tilde{\lambda}}&\le\left(\prod_{1\le i<j< d}(\lambda_i-\lambda_j-i+j)\right)\left(\prod_{1\le l<d}(\lambda_l-\lambda_d-l+d)\right)\\
&\le C_{{\max},d}\left(c_{\max,d}\cdot n\right)^{\frac{d(d-1)}{2}}.
\end{align}
Here
\begin{align}
C_{{\max},d}&:=\left(\prod_{1\le i<j< d}\left(j-i+1+\frac{3(d-1)}{c_{\max,d}\cdot n}\right)\right)\left(\prod_{1\le l<d}\left(2d-l-1+\frac{2(d-1)}{c_{\max,d}\cdot n}\right)\right)\\
&\le \left(2(d-1)+\frac{3}{c_{\max,d}\cdot n}\right)^{\frac{d(d-1)}{2}}.
\end{align}
Since $|\set{S}_{\rm Young}|=N^{d-1}$, we have
\begin{align}
d_{\rm P}&=\sum_{\lambda\in\set{S}_{\rm Young}}d_\lambda^2\\
&\le \left(2(d-1)+\frac{3}{c_{\max,d}\cdot n}\right)^{d(d-1)}\left(c_{\max,d}\, n\right)^{d^2-1}\\
&\le \left(2(d-1)c_{\max,d}\cdot n+3\right)^{d^2-1}.
\end{align}

\qed

\end{widetext}

\end{document}